# Directed molecular transport in an oscillating symmetric channel

D. Lichtenberg[a], A.E. Filippov[b] and M. Urbakh[a]

[a]*School of Chemistry, Tel Aviv University, 69978 Tel Aviv, Israel.*

[b]*Donetsk Institute for Physics and Engineering of NASU, 83144, Donetsk, Ukraine,*

**A mechanism responsible for the directed transport and molecular separation in a *symmetric* channel is proposed. We found that under the action of spatial harmonic oscillations of the channel, the system exhibits a directed transport in either direction, presenting multiple current reversals as the amplitude and/or frequency of the oscillations are varied. The particles of different masses may be forced to move with different velocities in the same or in the opposite directions by properly adjusting driving parameters. The directed transport can be produced in both directions even in the absence of thermal noise, the latter can speed up or slow down the transport depending on the system parameters.**

## I. Introduction

In recent years there has been an increasing interest in studies of pumping of ions, molecules and colloids through micro- and nanoscale channels [1-7]. These investigations have been motivated by a desire to understand how biological molecular pumps operate and to develop new strategies for a fabrication of synthetic pumps. Most studies in this field rely on the idea of particles moving in spatially asymmetric (ratchet) potentials, which can overcome simple diffusion under non-equilibrium conditions and gain directionality [8-12]. So far such studies have been focused on directed motion of particles through a rigid channel without considering fluctuations of its structure. However, investigations of transmembrane channels indicate that the dynamics of internal degrees of freedom of the channel proteins is in many cases essential for net directional motion to occur [13]. From the viewpoint of physics, it should also be interesting to explore the interconnections between the channel's intrinsic dynamics and directed transport. The closely related concept of *dynamical control* of motion has received much attention recently in the context of molecular engines [14-16]. It has been shown that directed motion can be induced dynamically and no static asymmetry is required which is built into in the system.

In this paper we propose a pumping mechanism driven by spatial fluctuations of a *symmetric* channel. We demonstrate that an effective pumping and separation of particles can be achieved by correlated oscillations of the walls of the channel in the lateral (along the channel axis) and normal (perpendicular to the channel axis) directions. The space oscillations of the channel lead to a modulation of the particle-wall interactions those produce temporally asymmetric forces acting on the particles. The main advantages of the proposed mechanism of pumping are: (a) the directionality of motion is determined dynamically and does not require any spatial asymmetry of the channel, (b) the transport velocity can be varied in a wide range, independent of the direction, (c) the pump allows a separation of particles according to their masses or/and interaction with the channel walls, (d) the pump can induce motion uphill a gradient of electrochemical potential, (e) depending on the parameters of the system, thermal noise can speed up or slow down the directed transport.

In the present paper we focus on investigations of particle motion in the underdamped regime. However, we demonstrate that the proposed mechanism of transport preserves its main features also in the overdamped regime that is most relevant for biological applications. It should be noted that in the latter case a separation of particles can be preformed by exploring the difference in their friction coefficients and interactions with the walls rather than using differences in the particle masses.

Inspired by a flexibility of biological systems we consider here a transport induced by spatial fluctuations of the channel walls. However, the same mechanism of directed motion and pumping can be realized in immobile channels, where similar variations of the particle-wall interactions can be achieved through an electric driving. For instance, the amplitude and the phase of symmetric space-periodic interaction between charged particles and walls can be modulated by an external ac electric field [17].

## II. The model

In order to demonstrate the characteristic properties of directed motion induced by the spatial fluctuations of the channel, we introduce a model of a particle embedded between two walls which oscillate in both the normal and the lateral directions (see Fig.1). The dynamical behavior of the particle is described by the two-dimensional Langevin equation of motion [18]:

$$m\ddot{\mathbf{r}} = -\eta\dot{\mathbf{r}} - \nabla_{\mathbf{r}}U(\mathbf{r}) + \mathbf{f}(t) \qquad (1)$$

Here $m$ and $\mathbf{r}=(x,y)$ are the mass and coordinate of the particles, and $\eta$ is the friction coefficient. The effect of thermal motion on the embedded particles is given by a random force, $f(t)$, which is $\delta$-correlated, $<f_i(t)f_j(0)> = 2\eta k_B T \delta(t)\delta_{ij}$. Here $T$ is the temperature and $k_B$ is the Boltzmann constant. The interaction between the particle and the channel is represented by the two dimensional potential $U(\mathbf{r})$:

$$U(\mathbf{r}) = U_0[\cos(\frac{2\pi(x+b(t))}{l}) + \sigma][\exp(\frac{y-a(t)/2}{\lambda}) + \exp(\frac{a(t)/2-y}{\lambda})] \quad (2)$$

which is periodic along the channel (x- direction) and presents a repulsion from the channel walls (y- direction). Here $l$ is the periodicity in x-direction, $a$ is the width of the channel and $\lambda$ is the characteristic length of particle-wall interaction in y-direction.

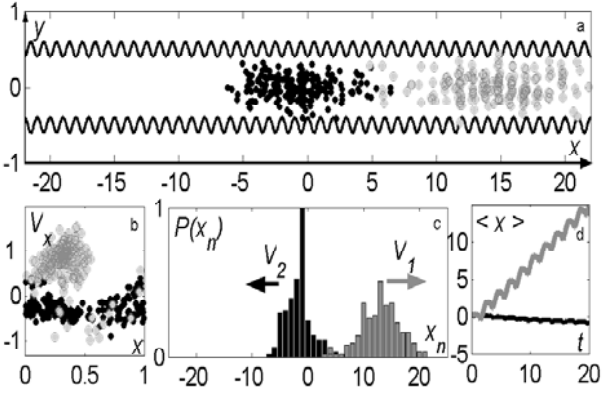

**Figure 1.** Directed transport and particle separation in a symmetric channel: (a) a two-dimensional snapshot of the embedded system obtained after ten periods of oscillations, (b) a snapshot of the system in the phase space $\{x_i, \partial x_i/\partial t\}$, (c) a spatial distribution of particles, $P(x_n)$, along the channel axis, (d) time dependences of the ensemble-averaged displacement of particles, $<x(t)>$. The results for particles of masses $m_1=1$ and $m_2=2.5$ are shown by gray and black colors correspondingly. Parameter values: $\omega/\omega_0(m_1)=1$, $\lambda/a_0=0.5$, $\Delta a/a_0=0.85$, $\Delta b/l=1.3$, $k_BT/U_0=0.02$ and $\eta/(m\omega_0(m_1))=0.89$.

The possibility of fluctuations of the channel width and of the lateral position of the walls is taken into account by introducing a time dependence into the channel width, $a(t)$, and the phase of periodic potential, $b(t)$. To keep the discussion simple, we will assume that both lateral and normal fluctuations follow a harmonic law, $b = \Delta b \cos(\omega t)$ and $a = a_0 + \Delta a \cos(\omega t)$, with the same frequency, $\omega$, and with a zero phase shift between them. In this case the expansion (opening) of the channel is accompanied by lateral displacement of the walls to the right, while the narrowing (closing) of the channel occurs with the displacement to the left. The space oscillations of the channel lead to a modulation of the particle-wall interaction: the amplitude of the periodic potential, $U(x, y = const)$, peaks at the minimum width of the channel (the close state) and goes through a minimum at the maximum width (the open state). The time average of the external force during this process is zero.

In contrast to ratchet models where the directed motion is determined by a spatial asymmetry of underlying potential, here we consider a system that is symmetric in space, and the symmetry is broken dynamically. The asymmetry in time is induced by coupling between normal and lateral oscillations that produces temporally asymmetric forces acting on the particles. It should be noted that neither lateral nor normal oscillations of the wall alone lead to a directed motion.

In order to characterize pumping we introduce an average current as,

$$J = \lim_{t \to \infty} \frac{1}{Nt} \sum_{j=1}^{N} \int_0^t \dot{x}_j(t')dt' \quad (3)$$

Calculations of current involve the averaging over the time and $N$ realizations. The definition (3) ensures the uniqueness of the current for a given frequency and amplitude of modulation. To understand the nature of the directed transport through the channel we also follow a time-dependent displacement of particles averaged over the ensemble of realizations, $<x(t)> = \frac{1}{N}\sum_{j=1}^{N} x_j(t)$.

The dynamical behavior of the system is determined by the following dimensionless parameters: $k_BT/U_0$, $\Delta b/l$, $\Delta a/a_0$ and $\lambda/a_0$, $\omega/\omega_0$, $\eta/(m\omega_0)$, where $\omega_0 = \sqrt{(2\pi U_0/lm)}$ is the frequency of small oscillations of the particle around the minimum of the periodic potential $U(x)$. The result presented in Figs.1-4 and Figs.6-8 are obtained for underdamped conditions, $\eta/(m\omega_0) < 1$, while the results in Fig.5 correspond to overdamped motion, $\eta/(m\omega_0) \gg 1$.

### III. Result and Discussion

Before analyzing the model and describing different regimes of motion, we present in Fig.1 a typical example of the time evolution of an ensemble of noninteracting particles driven by spatial oscillations of the symmetric channel. At the initial moment 200 particles of mass $m_1=1$ and 200 particles of mass $m_2=2.5$ have been placed around the potential minimum at the point $r=(0,0)$. The results for particles of masses $m_1=1$ and $m_2=2.5$ are shown by gray and black colors correspondingly. Fig.1 demonstrates the main features of the model: (a) the spatial oscillations of the channel lead to directional motion of the embedded particles both to the right and to the left, (b) both direction and velocity of motion depend not only on the driving parameters, $\omega$, $\Delta b$, but also on the particle masses that allows to use this mechanism for separation and mixing purposes. It should be noted that here we observe the directed transport in a *symmetric* potential under harmonic modulation of particle-wall interactions.

The characteristic property of the proposed mechanism of directed motion is that the time-averaged particle velocity is larger than the velocity of growing of the width of the particle spatial distribution (see Fig.1). This allows an effective manipulation by the ensemble of particles that is impossible in most ratchet systems where both velocities are of the same order [10].

The snapshots in the phase space (Fig.1b) and the time dependence of the particle displacement (Fig.1d) demonstrate that there is a qualitative difference between the motion of particles to the right and to the left that reflects an asymmetry in the channel oscillations. The motion to the right is more effective since the displacement of the channel wall to the right is accompanied by a decrease of the amplitude of the particle-wall potential.

In order to clarify the nature of the directed motion and current reversal we show in Fig.2 the frequency dependence of the ensemble-averaged particle displacement, $\Delta x_{op,cl}$, during the half-periods of oscillations which correspond to the opening and closing of the channel. For low frequencies of oscillations, $\omega \ll \omega_0$, the particle follows the oscillations of the minimum of the potential $U$ and there is no net displacement in this case. With increase in $\omega$, the particle has no time to respond to the driving force when the channel is open, and it leaves the minimum of the potential when the channel starts to narrow. As a result, for $0.4 < \omega/\omega_0 < 0.6$ the displacement to the left decreases while the displacement to the right is kept nearly constant. The net result is the transport to the right. This effect is most pronounced for the deterministic case ($T=0$) where we found a stepwise decrease of the displacement to the left and a corresponding increase of the net displacement as $\omega$ increases (Fig.2a). Analysis of the bifurcation diagram shows that the jumps in the displacements observed for $0.4\omega_0 < \omega < 0.6\omega_0$ correspond to transitions between different periodic orbits in the phase space. In this region of frequencies the averaged current can be written as $J = nl\omega/(2\pi)$, where $n=0,1,2,\ldots$, and for the parameters used in Fig.2a the maximal current equals $J = l\omega/\pi$. We remark that for $\omega < 0.6\omega_0$ particles perform fast asymmetric jumps to the right and to the left spending most of the time near some minimum of the potential $U$.

Further increase of frequency, $0.6 < \omega/\omega_0 < 0.8$, leads to a transition from periodic to chaotic motion of particles that occurs through a period-doubling route. In this regime we observe a steep decrease of the mean velocity followed by the current reversal at $\omega \approx 0.7\omega_0$. The current levels down to $J = -l\omega/\pi$ at the frequency $\omega/\omega_0 \approx 0.8$ at which a reverse bifurcation from chaotic to periodic orbits takes place. In contrast to the positive current arising from a difference in displacements to the right and left both being of the order of $l$, the negative current for $\omega/\omega_0 \approx 0.8$ is completely dominated by the particle motion to the left. The latter starts when the channel approaches the close state and ends in the open state. The snapshot in the phase and the time dependence of the particle displacement, which are presented in Figs.1b and 1d, reflect the above-mentioned features of the dynamics.

Fig.2b shows that thermal fluctuations smear out the stepwise variation of the particle displacement with frequency which is clearly visible in Fig.2a for $T=0$. However, as long as the thermal energy $k_BT$ is much smaller than the energy scale $U_0$ and the main features of the $\langle x(t,\omega) \rangle$ remain unchanged.

A contour map for the current in $(\omega, \Delta b)$-space (Fig.3) demonstrates that the proposed mechanism of pumping can produce a directed transport in either direction and gives rise to multiple current reversals as the frequency, $\omega$, and/or amplitude of the lateral oscillations, $\Delta b$, are varied. Here the amplitude of normal oscillations and temperature were kept constant. Regions of positive and negative current are displayed by light and dark colors correspondingly, and the curves separating these regions represent lines of constant current. One clearly sees three different domains in the $\{\omega, \Delta b\}$-space: (a) a region of small $\omega$ and $\Delta b$ characterized by a zero current, (b) alternating direction of constant $\Delta b$, and (c) a strip of high positive current separating the above-listed domains.

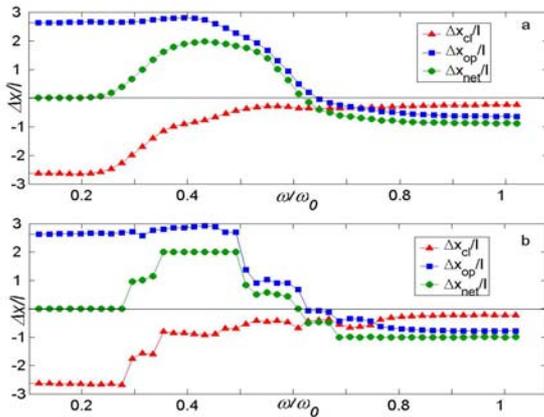

**Figure 2.** Frequency dependence of the ensemble-averaged displacement, $\Delta x_{op,cl}$, during the half-periods of oscillations, which correspond to the opening and closing of the channel, and of the net particle displacement, $\Delta x_{net}$, during the period of oscillations: (a) $T=0$, (b) $k_BT/U_0=0.02$. The number of realizations is $N=400$, other parameter as in Fig.1.

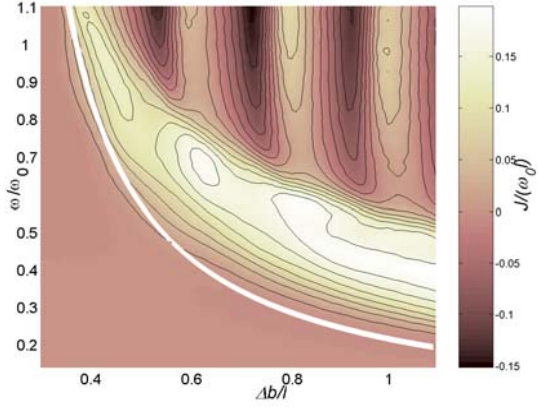

**Figure 3.** A contour map for the current, $J$, in the $\{\omega, \Delta b\}$-space. Regions of positive and negative current are displayed by light and dark colors respectively; the bar to the right of the map sets up a correspondence between the colors and the values of dimensionless current, $J/(l\omega_0)$. The white curve presents analytical estimations for $\omega_{cr}$ vs $\Delta b_{cr}$, $k_B T/U_0 = 0.02$ other parameter values as in Fig.1.

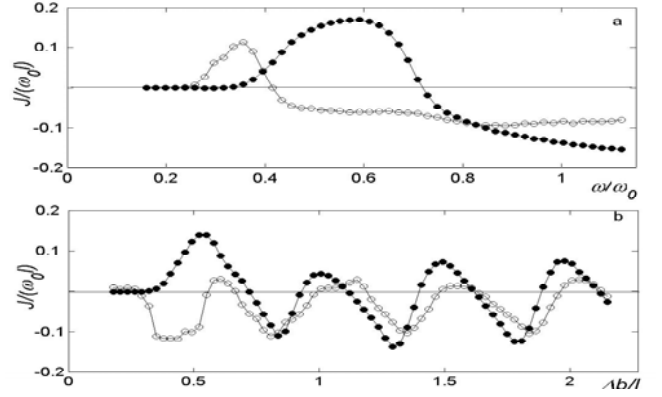

**Figure 4.** A current as a function of frequency (a), and amplitude (b) of lateral oscillations for particles of masses $m_1=1$ (closed circles) and $m_2=2.5$ (open circles). The frequency $\omega$ is normalized to the characteristic frequency $\omega_0$ calculated for $m=m_1=1$. Parameter values as in Fig.3.

The critical values of $\omega$ and $\Delta b$ for which the directed transport becomes possible (a bottom boundary of the strip (c)) can be estimated calculating a minimal particle velocity needed to jump to the neighboring minimum of the potential $U$ at the turning point where the velocity of the walls equals zero. That gives an equation $(\omega_{cr}/\omega_0)(\Delta b_{cr}/l) \approx const$, where the constant is of the order of one. The white curve $\omega_{cr}$ vs $\Delta b_{cr}$ corresponding to this equation is shown in Fig.3, and it fits well the results of numerical calculations. The contour map shows that the directed transport can be produced in a wide range of driving frequencies and amplitudes. The desirable current can be achieved by tuning the parameters of the oscillatory drive.

In addition to ability to control the direction and magnitude of the current, the proposed mechanism of pumping allows to separate particles according to their masses. In order to demonstrate a selectivity of the directed transport to the particle mass, we present in Fig. 4a-b the dependence of the current on the frequency and amplitude of oscillations which have been obtained for the particles of different masses. The results for $m_1=1$ and $m_2=2.5$ are shown by closed and open circles respectively. The most pronounced effect of mass on the transport comes through the $m$-dependence of the characteristic frequency, $\omega_0 \propto m^{-1/2}$, that determines the scale of variation of $J$ with $\omega$. This explains the shifts of frequencies corresponding to the maximum, $\omega_{max}$, and to the point of current reversal, $\omega_{rev}$, in the $J(\omega)$-curves in Fig.4a, which are caused by the change of particle mass, namely, $\omega_{max,rev}(m_1)/\omega_{max,rev}(m_2) \approx (m_2/m_1)^{1/2}$.

There is also a considerable distinction between $J(b)$ curves obtained for different masses (see Fig.4b). We note that the current as a function of the amplitude $\Delta b$ exhibits multiple peaks that correspond to parametric resonances arising under the lateral oscillatory drive. The resonant bands in $J(\Delta b)$ are also clearly visible in the contour map in Fig.3.

Figures 4a-b demonstrate that by properly adjusting parameters $\omega$ and $b$, the particles of different masses may be forced to move with different average velocities in the same or in the opposite directions. Alternatively particles of a certain mass may be forced to stay localized while others move in a desirable direction with a desirable velocity. Thus, the proposed pump can indeed act as an extremely selective device for separating different types of particles. The pumping and separation can be optimized and further controlled by adjusting the driving parameters.

The proposed device opens interesting perspectives for manipulating reaction-diffusion systems, e.g. to mix reactants with different masses and to provide a transport of products into the desirable direction. In contrast to many conventional techniques of separation, such as centrifugation and chromatography that rely on long-range gradients, we propose a mechanism of separation, which can be realized at micro and nanoscales and does not require a presence of global gradients.

In the present paper we focus on directed transport in the underdamped regime, $\eta/(m\omega_0) < 1$. However the proposed mechanism of pumping allows to produce a directed motion also under overdamped conditions, $\eta/(m\omega_0) \gg 1$. A typical frequency dependence of the current obtained in this case is shown in Fig.5. We found that the maximal value of current in the overdamped case is smaller than that for the underdamped conditions. However, the parametric resonances are more pronounce in the overdamped case than for undredamped conditions, and as a result the value of the current can be easily controlled by changing the driving frequency and/or amplitude of the oscillations. In the overdamped case the particles cannot be separated according their masses however the separation is still possible, e.g. according to their friction coefficient, $\eta$, and interaction with the walls. In contrast to the underdamped case here the direction of the motion is dictated by an asymmetry of the drive and does not depend of the frequency and the amplitude of modulation. In the absence of a phase shift between oscillations in the normal and lateral directions the current is positive but its direction can be reversed by including a phase shift.

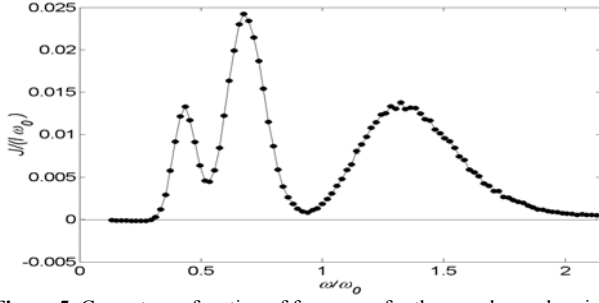

**Figure 5**. Current as a function of frequency for the overdamped regime, $m=0$, $\Delta b =1.7$ other parameter values as in Fig.3.

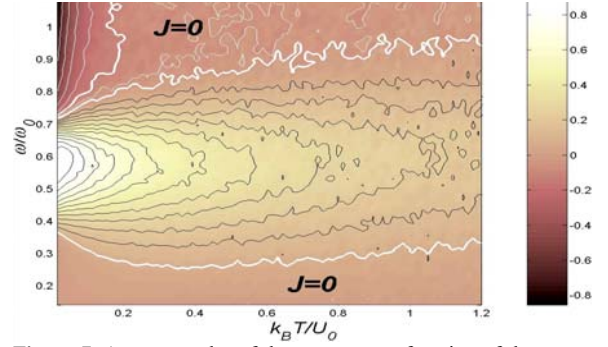

**Figure 7.** A contour plot of the current as a function of the temperature and of the frequency. Regions of positive and negative current are displayed by light and dark colors respectively; the bar to the right of the map sets up a correspondence between the colors and the values of dimensionless current, $J/(l\omega_0)$.

The proposed pump is powerful enough to produce a transport of particles against a gradient of electro-chemical potential. We simulate the effect of electro-chemical potential by applying a constant force at each particle. The maximum force, $F_m$, that the pump may resist and still produce a directed motion, can serve as a measure of how powerful the device is. Our calculations show that this force is of the order of $U_0/l$, and it depends sensitively on driving parameters, $\omega$, $\Delta b$ and temperature. A frequency dependence of the maximum force is shown in Fig.6. For higher forces, the particles first remain in their initial location, and finally move with the force. It should be noted that in the absence of the external force the current reversal has been observed as the frequency changes (see Fig.4 ). Accordingly the maximum resisting force presented in Fig.6 changes it sign at $\omega = 0.71$.

To complete the discussion of transport properties we consider the effect of temperature on the current. Fig.7 presents a contour map for the current as a function of temperature and driving frequency. As in Fig.3 regions of positive and negative current are displayed by light and dark. One sees that high absolute values of the current have been observed at low temperatures, $k_BT/U_0<0.3$. Both negative and positive current can be produced in this region. In the high temperature region, $k_BT/U_0>1$, the magnitude of the current decreases and only the positive current is possible.

In Fig.8 we show three representative examples of temperature dependence of the current. Our results allow to conclude that: (a) the directed transport can be produced in both directions even in the absence of thermal noise, (b) thermal noise can induce a current that is absent otherwise, (c) depending on the parameters of the system thermal noise can speed up or slow down the directed transport, and (d) one can also find pure noise-induce current reversals from a negative to positive current, and

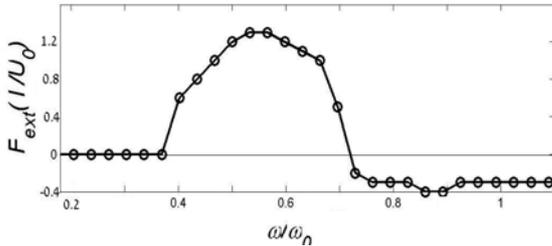

**Figure 6.** Frequency dependence of the dimensionless maximum force, that the pump can resist and still produce a directed motion. Parameter values as in Fig.3.

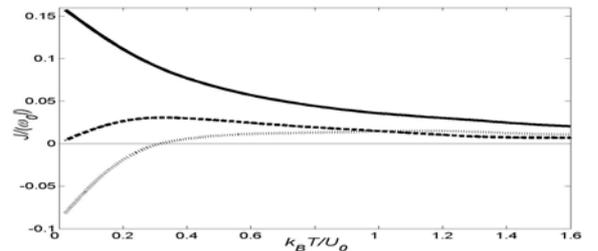

**Figure 8.** Temperature dependences of the current calculated for three values of dimensionless frequency, $\omega/\omega_0$=0.56 (solid line), 0.36 (dashed line) and 0.7 (dotted line). Parameter values as in Fig.3.

(e) as aforementioned, in the high temperature region, $k_B T > U_0$, only the positive current is possible.

## IV. Conclusions

We have found that effective molecular pumping and separation can be produced by spatial oscillations of walls of a symmetric channel. The system exhibits a directed transport in both directions, presenting multiple current reversals as the amplitude and/or frequency of the oscillations are varied. In the underdamped regime the pump allows a separation of particles according to their masses, and/or friction coefficient and interaction with the walls. The optimal pumping and separation can be achieved by tuning the driving parameters in with the walls. The optimal pumping and separation can be achieved by tuning the driving parameters in a controllable way. Depending on the parameters, the addition of thermal noise can speed up or slow down the particle motion and induce a current reversal that is absence otherwise. The proposed mechanism of pumping allows also to produce and tune a directed motion for overdamped conditions. In this case particle separation can be preformed according to their friction coefficient and interactions with the walls of the channel.

**Acknowledgment**

Financial support for this work by grants from the Israel Science foundation (Grant No 573/00) and BSF is gratefully acknowledged. A.E.F thanks ESF Scientific Program "Nanotribology" for financial support.